\documentstyle[prl,twocolumn,aps,epsf]{revtex}
\begin{document}
\draft

\title{
Spin flip scattering in magnetic junctions}
\author{F. Guinea}
\address{
Instituto de Ciencia de Materiales.
Consejo Superior de Investigaciones Cient{\'\i}ficas.
Cantoblanco. 28049 Madrid. Spain.  \\}
\date{\today}
\maketitle 
\begin{abstract}
Processes which flip the spin of an electron tunneling in 
a junction made up of magnetic electrodes are studied.
It is found that: i) Magnetic impurities give a contribution 
which increases the
resistance and lowers the magnetoresistance, which saturates at low
temperatures. The conductance increases at high fields.
ii) Magnon assisted tunneling reduces the magnetoresistance
as $T^{3/2}$, and leads to a non ohmic contribution to the
resistance which goes as $V^{3/2}$, iii) Surface
antiferromagnetic magnons, which may appear if the interface
has different magnetic properties from the bulk, gives rise
to $T^2$ and $V^2$ contributions to the magnetoresistance
and resistance, respectively, and, iv) Coulomb blockade effects
may enhance the magnetoresistance, when transport is dominated
by cotunneling processes.
\end{abstract}


\narrowtext
\section{Introduction}

Magnetic junctions made up of fully saturated ferromagnets
have attracted a great deal of attention, as they may lead to
large magnetoresistance at low fields.
This effect should be particularly enhanced in fully polarized
magnets\cite{WK55,CVM97}. 
The tunneling probability between two fully polarized
electrodes whose magnetization are at a relative angle
$\theta$ goes like $\cos^2 ( \frac{\theta}{2} )$. By averaging this
quantity, we find that the conductance in the absence of
an applied field, when $\theta$ can have any value, is
one half of the conductance in the presence of a field,
which aligns the magnetization and makes $\theta = 0$.
This simple analysis predicts that the maximum allowed
magnetoresistance is 100\%.
This upper bound remains to be achieved in experiments,
although enhanced effects have been reported in different 
experimental setups\cite{HCOB96,Letal96,Getal96,Vetal97}. 
Spin tunneling is also expected to be dominant in transport
through ceramics and granular systems, where enhanced magnetoresistance
at low fields has also been reported\cite{Setal96,Metal96,BFMO97,FMBO97}.
The relevance of magnetic scattering at the interface
in perovskite manganites can also be inferred
by comparing with transport in related materials 
which exhibit colossal magnetoresistance\cite{HC97}.

In the present work, we study various effects which may
limit the observed magnetoresistance.
The bound discussed above assumes that
the transmitted electron has a well defined spin throughout
the tunneling process. If the spin can flip as the electron
hops from one electrode to the other, the observed magnetoresistance
will be reduced with respect to the previous value.
Thus,  we shall consider processes which change the spin 
of the electron as it tunnels.

In the following section, we analyze the role of magnetic
impurities which may be present in the interface region.
For simplicity, we will assume that they are magnetic ions
of the same type as those which exist in the bulk of
the electrodes, that is, Mn or Cr.

The third section studies spin flip processes which involve the 
excitation of bulk magnons during
the tunneling process.

In the fourth section, we consider the possibility that the
interface has different magnetic 
properties than the bulk\cite{Vetal97,FMBO97,XCN87,P93}.
In particular, we analyze the influence of an antiferomagnetic
layer at the interface, which may be present if it is oxidized,
for instance.

The fifth section analyzes the main consequences of Coulomb blockade
on spin polarized tunneling. Coulomb blockade requires large charging 
energies, and small dimensions. It can be relevant to transport in
granular materials.

The main conclusions are presented in the last section.

\section{Magnetic impurities}
The current between the electrodes may arise from direct
electron transfer, or by processes in which electrons hop
into impurity levels within the barrier between the
electrodes. If the impurities are magnetic, these processes
can change the spin of the electron, and modify the
observed magnetoresistance. We will mostly assume that the contribution
of magnetic impurities to the total current, although at the end
of the present section some comments are made on the
features to be found in the opposite limit.

\begin{figure}
\begin{center}
\mbox{\epsfxsize 8cm\epsfbox{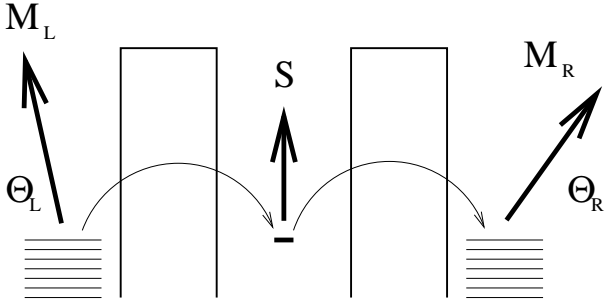}}
\end{center}
\caption{Sketch of the elastic spin flip processes mediated by
impurities at the interface. See text for detains.}
\label{imp}
\end{figure}

We consider impurities, such as Mn$^{3+}$, Mn$^{4+}$, Cr$^{3+}$
and Cr$^{4+}$ of the same kind as the ions present in the
electrodes. These impurities have electronic levels at energies
within the conduction band of the electrodes. Electrons, or holes,
can hop from one electrode into these levels, and from there
to the other electrode (see fig. (\ref{imp})). We assume that
the Hund's coupling between these electrodes and the core spins
is much larger than any other scale in the problem. Thus,
the hopping process only involves the highest spin state of
the ion. 

At zero temperature, an electron
(or hole) which is initially polarized in the direction of the 
magnetization of one electrode has a finite amplitude of hopping
into the impurity. In a basis aligned with the impurity spin,
the initial state of the impurity is $| S , S \rangle$. After
an electron hops from the left electrode, it becomes: 
\begin{equation}
\cos \left(  \frac{ \theta_{LI} }{2} \right) | S + \frac{1}{2} ,
S + \frac{1}{2} \rangle + \frac{
\sin \left( \frac{\theta_{LI}}{2} \right)}{
\sqrt{2 S + 1}} | S + \frac{1}{2} , S - \frac{1}{2} \rangle
\end{equation}
where  $\theta_{LI}$ is the relative angle between the magnetization
of the left electrode and the initial spin of the impurity.
We are projecting out intermediate states where the spin of
the impurity is not maximum.

Let us now assume that the same electron hops coherently into the right
electrode. The spin of the electron must be
parallel to the magnetization of the right electrode.
The direction of the magnetization forms an angle $\theta_{IR}$
with respect to the spin of the impurity, and an azimuthal angle
$\phi$ with respect to the plane formed by the impurity spin
and the magnetization of the right electrode.
Then, the final state of the impurity is:
\begin{eqnarray}
 & &\left[ \cos \left( \frac{\theta_{LI}}{2} \right)  
 \cos \left( \frac{\theta_{IR}}{2} \right)  + \frac{
 \sin \left( \frac{\theta_{LI}}{2} \right)  
 \sin \left( \frac{\theta_{IR}}{2} \right) e^{i \phi}}{2 S + 1} \right]
| S , S \rangle \nonumber \\ &+ &\frac{ \sqrt{2 S} 
 \sin \left( \frac{\theta_{LI}}{2} \right)  
 \cos \left( \frac{\theta_{IR}}{2} \right) }{2 S + 1}
| S , S - 1 \rangle
\end{eqnarray}
In the previous analysis we assumed that the impurity can accept
an electron, as for Mn$^{4+}$. A hole current, in the opposite
direction, can take place through ions like Mn$^{3+}$. The corresponding
calculation is straightforward, except that the angles are interchanged.

Finally, the probability that an electron hops between the
two electrodes after a process like the one described above is:
\begin{eqnarray}
{\cal T} &\propto  &{} \nonumber \\ &\langle &\left[   
\cos \left( \frac{\theta_{LI}}{2} \right)  
 \cos \left( \frac{\theta_{IR}}{2} \right)  + \frac{
 \sin \left( \frac{\theta_{LI}}{2} \right)  
 \sin \left( \frac{\theta_{IR}}{2} \right) e^{i \phi}}{2 S + 1} \right]^2
 \rangle \nonumber \\ &+ &\frac{2 S}{(2 S + 1)^2} \langle
 \sin^2 \left( \frac{\theta_{LI}}{2} \right)  
 \cos^2 \left( \frac{\theta_{IR}}{2} \right)  \rangle
\label{transmission}
\end{eqnarray}
where the brackets denote thermal averages over the values of the
angles $\theta_{LI}$, $\theta_{IR}$ and $\phi$. In the case that
the direction of the impurity spin is totally random, we
find that ${\cal T} \propto \frac{2 S^2 + 3 S + 1}{2 ( 2 S + 1 )^2}$.
For $S = \frac{3}{2}$, $T \propto \frac{5}{16}$. This value
increases in an applied field, as the impurities tend to be
aligned by it. The corrections can be obtained by performing
the averages in (\ref{transmission}) in the presence of a field.
Expanding, we find:
\begin{equation}
{\cal T} = \left\{ \begin{array}{l} \frac{2 S^2 + 3 S + 1}{2 
( 2 S + 1 )^2} + \frac{\mu_0 S H}{k_B T} \frac{S^2 + S}{4 ( 2 S + 1 )^2} +
\left( \frac{\mu_0 S H}{k_B T} \right)^2 
\frac{2 S^2 + S + 1}{8 ( 2 S + 1 )^2} + ...  \\
\frac{\mu_0 S H}{k_B T} \ll 1 \\
1 - \frac{k_B T}{\mu_0 S H} \frac{4 S^2 + 1}{2 ( 2 S + 1 )^2}
+ \left( \frac{k_B T}{\mu_0 S H} \right)^2 \frac{1}{( 2 S + 1 )^2}  + ...
 \\ \frac{\mu_0 S H}{k_B T} \gg 1 \end{array} \right.
\label{expansion}
\end{equation}

By comparing the tranmission with and without a field, we
find that the contribution of impurity scattering can be
included into the magnetoresistance as:
\begin{equation}
\frac{\Delta \sigma}{\sigma} = \frac{\sigma_0 ( H , T ) - 
\sigma_0 ( 0 , T ) + \sigma_I 
\frac{ \mu_0 S H }{k_B T } \frac{S^2 + S}{4 ( 2 S + 1 )^2}}
{ \sigma_0 ( H , T ) +
\sigma_I \left[ 
\frac{2 S^2 + 3 S + 1}{2 ( 2 S + 1 )^2} +
\frac{ \mu_0 S H }{k_B T } \frac{S^2 + S}{4 ( 2 S + 1 )^2} \right] }
\label{MRimp}
\end{equation}
This expression is valid
to lowest order in $\frac{\mu_0 S H}{k_B T}$.  In this expression,
$\sigma_0$ stands for all contributions to the conductance other than
those due to the impurities, and $\sigma_I$ is the conductance due to
impurities at zero field. 
The magnetoresistance, when scattering by magnetic impurities
is allowed, is reduced by a factor 
proportional to $\frac{\sigma_I}{\sigma_0}$, and it is
temperature independent at low fields.
As these processes are elastic, they do not induce a dependence on 
applied voltage.
The crossover between the low and the high field regime 
takes place when $\frac{\mu_0 S H}{k_B T} \sim 1$. If $S = \frac{3}{2}$
and $T = 300$K, this field is 60T, while for $T = 4$K, the field is 
0.8T. The corresponding figures for $S = \frac{1}{2}$ impurities are
180T and 2.4T.

Finally, if most of the current was due to resonant tunneling through
magnetic impurities, the observed magnetoresistance will
increase, instead of decreasing as in the previous case.
A magnetic filed will align the spin of the impurities with
the magnetization of the electrodes.
The impurities behave in a similar way to a third magnetic layer
located between the electrodes. 
For instance, the current which flows from tunneling through 
$S = \frac{3}{2}$ impurity levels whose magnetic moments 
are oriented at random 
is $\frac{5}{16}$ that of the current
when the moments are aligned by a field.
The corresponding figure for tunneling between magnetic electrodes
is $\langle \cos^2 ( \frac{\theta}{2} ) \rangle = \frac{1}{2}$.
A possible situation where of the current in a magnetic
junction is due to sequential
(not resonant) tunneling through impurity levels is reported in\cite{DR97}.
\section{Spin flip processes induced by bulk magnons.}
The spin of the tunneling electron can be changed by the creation, or
absorption, of magnons in the electrodes. These processes increase
the tunneling probability between electrodes whose magnetization
is not aligned,
and reduce the observed magnetoresistance. 
We will consider that the rate of magnon induced tunneling is independent
of the relative angle between the magnetization in each electrode.

If the barrier width if of length $d$, the electron, after a tunneling
event, will be spread in a region of size $d$ in the electrode it has
hopped to. We write the electron creation 
operator as $\int f ( r ) \psi^\dag ( r )
b^\dag_s ( r )$, in terms of a spinless fermion, $\psi$, and a
Schwinger boson of spin $s$, $b$\cite{AA88}. The function $f ( r )$ 
gives the spatial extent of the wavefunction of the
electron after the tunneling process, $\sim d$, the thickness
of the barrier.
By expanding these operators into
the normal modes of the electrodes, we find that, roughly, all
magnons, $b_{\vec{k} s}$ with wavelengths larger than $d$
can be created with equal probability.

At zero temperature only magnon creation is allowed. This
is possible at finite junction voltages. An electron 
with energy $\epsilon$ above the chemical potential of the
other electrode can create any magnon with energy
$\epsilon ' < \epsilon$, provided that the wavelength
of the magnon, $a \left( \frac{\epsilon '}{J} \right)^{1/2}$
is less than $d$ ($a$ is the lattice constant).
The density of states of magnons
in a ferromagnetic three dimensional system is
$D ( \epsilon ' ) \propto \frac{1}{J} \left( \frac{ \epsilon '}
{ J } \right)^{1 / 2}$. Hence, the intensity due
to magnon creation is:
\begin{eqnarray}
I ( V ) &\sim &\frac{1}{R} \int_0^V d \epsilon \int_0^{
{\rm min} [ \epsilon , J ( a/d )^2 ]}
d \epsilon ' \frac{1}{J} \left( \frac{\epsilon '}{J} \right)^{
\frac{1}{2}}  \nonumber \\ &\sim 
&\left\{ \begin{array}{lr}
\frac{V}{R} \left( \frac{V}{J} \right)^{\frac{3}{2}} 
&V \ll J \frac{a^2}{d^2} \\
\frac{V}{R} \left( \frac{a}{d} \right)^3   &V \gg J \frac{a^2}{d^2}
\end{array} \right.
\label{magnonv} 
\end{eqnarray}
This contribution is to be added to the elastic conductance,
$I_0 ( V ) =  \frac{V}{R} \langle \cos^2 \left( \theta / 2 \right ) \rangle$.
We assume that magnon induced tunneling is independent of the
relative angle between the magnetization in the two electrodes.
Hence, the term shown in (\ref{magnonv}) reduces the magnetoresistance
at finite voltages.

At finite temperatures, the electrons which tunnel can excite magnons
of energy below $k_B T$. Using the previous argument, the probability
that an electron excites a magnon goes as $\left( \frac{k_B T}{J} \right)^{
\frac{3}{2}}$. Thus, the differential conductivity at low voltages
has a contribution which goes as:
\begin{equation}
\delta \sigma_{mag} \sim 
\left\{ \begin{array}{lr} \frac{1}{R} \left( \frac{k_B T}{J} \right)^{
\frac{3}{2}} &k_B T \ll J \frac{a^2}{d^2}  \\
\frac{1}{R} \left( \frac{a}{d} \right)^3 &k_B T \gg J \frac{a^2}{d^2}
\end{array} \right.
\label{conducmag1}
\end{equation}
This term increases the conductance of the junction, and it is
independent of the relative orientation of the magnetization of
the electrodes. Hence, the observed magnetoresistance decreases,
as $T$ increases, as $\left( \frac{k_B T}{J} \right)^{\frac{3}{2}}$.
This contribution has the same temperature dependence as the reduction of the
magnetoresistance due to the decrease in the magnetization of
the electrodes. The latter effect, however, does not give rise
to non linear I-V characteristics.
\section{Spin flip processes due to magnons at the interface.}
It is likely that, in doped manganites or in CrO$_2$, the surface has
a different composition than the bulk. 
In addition, the double exchange mechanism 
is weaker at a surface, as the kinetic
energy of the carriers is reduced. Both effects may reduce the 
tendency towards ferromagnetism, leading to antiferromagnetic
behavior. Note that a change in the magnetic structure of the
surface leads to modifications in the height of the tunneling 
barrier. It is, however, unlikely that a simple dependence of the height of
the barrier on the magnetic surface energy can be found\cite{Zetal97}.

The contribution of spin flip processes due to interface antiferromagnons
can be estimated in the same way as in the preceding section.
The only difference is the change in the density of states,
due to the different dispersion relation, and to the low
dimensionality. For two dimensional antiferromagnons,
this quantity is $D ( \epsilon ) \propto \frac{1}{J_{AF}} 
\left( \frac{\epsilon}
{J_{AF}} \right)$. The highest energy plasmon
which can couple to the tunneling electron has energy
$ \sim J_{AF} \frac{a}{d}$. Hence, the intensity depends on
voltage as:
\begin{equation}
I ( V ) \sim \left\{ \begin{array}{lr}
\frac{V}{R} \left( \frac{V}{J_{AF}}  \right)^2 &V \ll J_{AF} \frac{a}{d} \\
\frac{V}{R} \left( \frac{a}{d} \right)^2 &V \gg J_{AF} \frac {a}{d} \end{array}
\right.
\label{magnons} 
\end{equation}
and, at finite temperatures, we find a contribution to the conductivity
like:
\begin{equation}
\delta \sigma_{surf} \sim \left\{ \begin{array}{lr}
\frac{1}{R} \left( \frac{k_B T}{J_{AF}} \right)^2 &k_B T \ll J_{AF}
 \frac{a}{d} \\
\frac{1}{R} \left( \frac{a}{d} \right)^2 &k_B T \gg J_{AF} \frac{a}{d} 
\end{array} \right.
\label{conducmag2}
\end{equation}
As in the previous case, this effect reduces the observed magnetoresistance
at finite temperatures. If $a \sim d$ and $k_B T \gg J_{AF}$, the contribution 
of processes mediated by magnons is comparable to the purely elastic 
conductance. In this limit, the magnetoresistance should tend to zero.

It has been argued that, in tunnel junctions based on Co, ferromagnetic
magnons are localized at the interface\cite{Zetal97}. The scheme used here
can be appplied to this case. By inserting the appropiate density
of states, we recover the results reported in\cite{Zetal97}.
\section{Coulomb blockade effects.}
\begin{figure}
\begin{center}
\mbox{\epsfxsize 8cm\epsfbox{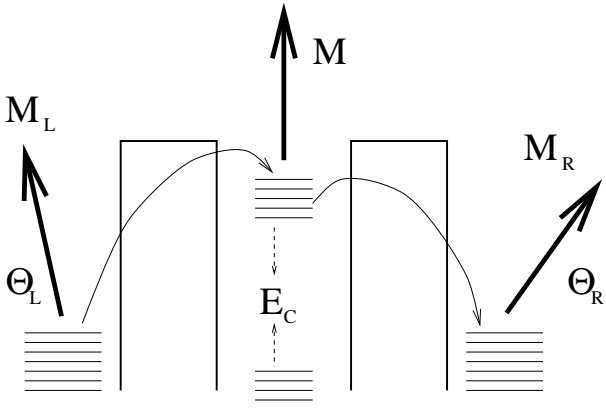}}
\end{center}
\caption{Sketch of the cotunneling process through a grain
with charging energy $E_C$.}
\label{cotunn}
\end{figure}

Coulomb blockade reduces the conductance of granular systems at low 
temperatures\cite{KS75,AL91}. 
The charging energy required to add one electron to a
grain, $E_C = \frac{e^2}{C}$, where $C$ is the capacitance of the grain,
is not negligible. It 
tends to open a gap when $k_B T \ll E_C$. The main process which 
suppresses this gap, at
low temperatures, is inelastic cotunneling\cite{AN92}. 
At finite temperatures or
voltages, one electron can hop into a grain and leave it on a time
scale shorter than $\hbar E_C^{-1}$, leaving an excited electron-hole
pair of energy $\epsilon < k_B T , V$. 
A sketch of the process is depicted in fig.(\ref{cotunn}).
Cotunneling gives a conductance
which goes as 
$\frac{\hbar}{e^2 R^2} \left( \frac{k_B T}{E_C} \right)^2$.
This estimate is valid when a single small grain inserted
between much larger grains blocks the current. If we consider $N$
grains in series, the conductance due to cotunneling goes like
$\frac{1}{R} \left( 
\frac{\hbar}{e^2 R} \right)^N  \left( \frac{k_B T}{E_C} \right)^{N+1}$.
A sketch of the process is depicted in fig.(\ref{cotunn}).

Cotunneling requires two correlated hopping processes. In a fully polarized
magnet, each hopping is reduced by a factor proportional to $\cos^2
( \frac{\theta}{2} )$, where $\theta$ is the angle between the 
magnetization in thea central
grain and that in the right or the left grains. Averaging over orientations, 
cotunneling is reduced by a factor $\frac{1}{2} \times \frac{1}{2} =
\frac{1}{4}$, when the magnetizations are at random.
The corresponding value for $N$ grains of small capacitance
is $\frac{1}{2^{N+1}}$.
This factor becomes 1 when a magnetic field aligns the
magnetization of the grains.
Hence, cotunneling is more sensitive to magnetic disorder that
direct tunneling. The magnetoresistance should increase 
when Coulomb blockade supresses direct tunneling, and only
cotunneling is allowed. Note that this effect is 
only important at very low temperatures, when activated
processes over the Coulomb barrier are negligible.
It should also be absent in magnetic junctions where the
electrodes are not fully polarized\cite{Setal97}.

Sufficiently small grains, however, will behave like 
non resonant (because of the Coulomb gap) magnetic impurities.
Hence, tunneling between misorented grains is not totally
suppressed, as discussed in section (II). This gives a limit
to the maximum achievable magnetoresistance.
In addition, the resistance of $N$ small grains in series
can be so large, $\sim ( 1 {\rm k} \Omega )^N$, that 
the magnetoresistance cannot be observed.

At intermediate temperatures, when the contribution from
cotunneling is small, the existence of a Coulomb gap
leads to activated transport. In the presence of magnetic disorder,
the activation energy also includes a contribution from
spin flip processes, which must take place during the
tunneling event\cite{HA76}.
This energy goes like the average magnon energy excited in the
tunneling process, as discussed in section (III), $E_M \sim J
\frac{a^2}{d^2}$. If we assume that all intergrain junctions are
identical, this effect leads to a magnetoresistance which
should increase as $e^{\frac{E_M}{k_B T}}$. This process, however,
is limited by cotunneling at low temperatures.
\section{Conclusions.}
Spin flip processes reduce the magnetoresistance of junctions between
fully polarized magnets. Their origin may be extrinsic, related to
the different magnetic properties of the interfaces, or intrinsic,
associated to the excitation of bulk magnons. 
In addition, they can be classified into elastic, as the scattering by
magnetic impurities, or inelastic, which are mediated by magnetic excitations.

Elastic spin flip processes can due to magnetic impurities 
or other static deviations
from perfect ferromagnetism,
like domain walls\cite{LZ97}. They are extrinsic, as they should not be present
in perfect systems. They give rise to a temperature independent reduction of
the magnetoresistance.  Assuming that the scattering by these imperfections
leads to a loss of the spin orientation of the electron, the relative
reduction in the magnetoresistance goes as $\frac{\sigma_I}{\sigma_0}$, where
$\sigma_I$ is the contribution to the conductance from
resonant tunneling via impurity states, and $\sigma_0$ stands for the 
conductance due to other tunnel processes.

Inelastic spin flip processes do not reduce the magnetoresistance at
zero temperature and zero voltage, but give rise to non ohmic effects
at finite voltages, and to changes in the
conductance as function of temperature.
We can distinguish between intrinsic effects, mediated by bulk magnons,
and those related to magnetic excitations of the interface.
Bulk magnons reduce the magnetoresistance at temperatures, or voltages,
comparable to the bulk exchange coupling, which is of the order
of the Curie temperature. The effect due to interface excitations
shows up at the scale of the new couplings at the interface.
In this work, we have considered the influence of an antiferromagnetic
layer at the interface, but more complicated excitations may exist 
if the surface is strongly disordered.

Finally, we have analyzed the interplay of spin polarized tunneling and
Coulomb blockade. We find that cotunneling processes enhance the
magnetoresistance. This effect may be difficult to observe,
due to the high resistance of junctions at the temperature when
Coulomb blockade is fully developed.
\section{Acknowledgements}
I am thankful to L. Brey, J. M. D. Coey, P. Fontcuberta and X. Obradors
for many helpful comments, suggestions and criticisms. This work
was supported by the CICyT (Spain) through grant PB96-0875.

\end{document}